\documentclass[letter,twocolumn]{jpsj2}

\def\runauthor{Katsunori \textsc{Kubo}}

\title{Fulde-Ferrell-Larkin-Ovchinnikov State
  in the Absence of a Magnetic Field}

\author{Katsunori \textsc{Kubo}}
\inst{Advanced Science Research Center,
Japan Atomic Energy Agency,
Tokai, Ibaraki 319-1195}

\recdate{January 28, 2008; 
  accepted February 8, 2008; published March 25, 2008}

\abst{
We propose that in a system with pocket Fermi surfaces,
a pairing state with a finite total momentum $\mib{q}_{\text{tot}}$
like the Fulde-Ferrell-Larkin-Ovchinnikov state can be stabilized
even without a magnetic field.
When a pair is composed of electrons on a pocket Fermi surface
whose center is not located at $\Gamma$ point,
the pair inevitably has finite $\mib{q}_{\text{tot}}$.
To investigate this possibility,
we consider a two-orbital model on a square lattice
that can realize pocket Fermi surfaces and we apply fluctuation exchange approximation.
Then, by changing  the electron number $n$ per site,
we indeed find that such superconducting states
with finite $\mib{q}_{\text{tot}}$ are stabilized
when the system has pocket Fermi surfaces.
}

\kword{FFLO state, pocket Fermi surfaces, orbital degree of freedom, fluctuation exchange approximation}

\begin{document}
\maketitle

The pairing states of superconductivity are classified
on the basis of symmetries that satisfy the Pauli principle~\cite{Sigrist}.
Among the pairing states,
the $s$-wave state is the simplest
and is called conventional superconductivity.
The other superconducting states are anisotropic (unconventional) ones.
We can categorize
the superconducting states into spin-singlet and spin-triplet states.
The spin-singlet states have even parity, e.g., $s$-wave,
under the inversion of space due to the Pauli principle.
The spin-triplet states have odd parity,
and in these states, the spin degree of freedom is active
and many phases can occur in a system.

If we can introduce an additional degree of freedom into a system,
we can find exotic superconducting states that are beyond
the ordinary categorization.
For this purpose,
the superconductivity in systems with an orbital degree of freedom,
which is not considered in the above discussion on superconducting symmetry,
has been studied recently.
For example, the effects of the orbital degree of freedom on superconductivity
have been studied for
a two-orbital Hubbard model with the same dispersion for both orbitals,
by a mean-field theory~\cite{Klejnberg},
dynamical mean-field theory~\cite{Han,Sakai},
and a fluctuation exchange (FLEX) approximation~\cite{Kubo_2OHM}.
These studies have revealed that
an $s$-wave spin-triplet state and a $p$-wave spin-singlet state,
which satisfy the Pauli principle
by composing the orbital state of a pair antisymmetrically,
can be stabilized in the two-orbital Hubbard model.

Although the two-orbital Hubbard model provides such an interesting possibility
of superconductivity in a multi-orbital system,
it describes a system with a rather high orbital symmetry,
that is, the Fermi surfaces of these orbitals are exactly the same.
To discuss more realistic situations,
we should improve the two-orbital Hubbard model.
In particular, we should include the effects of orbital anisotropy,
for example, orbital-dependent hopping integrals
and the transformation property of orbitals.
These properties are important for magnetism in $d$-electron systems~\cite{Imada, Hotta}
and $f$-electron systems~\cite{Hotta,Kubo_NpO2,Kubo_Gamma8},
and they should also be important for superconductivity.
Among the above-mentioned properties, the multi-Fermi-surface nature
due to the orbital-dependent hopping is important,
since a pair on a Fermi surface can interact
with a pair on another Fermi surface,
which may provide a new pairing mechanism.

\begin{figure}[t]
  \begin{center}
    \includegraphics[width=0.5\linewidth]{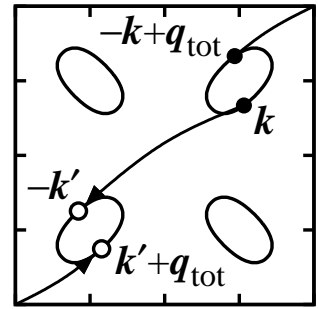}
  \end{center}
    \caption{\label{figure:pocket_FS}
    Pocket Fermi surfaces.
    A pair with a total momentum $\mib{q}_{\text{tot}}$
    is composed of electrons with $\mib{k}$ and $-\mib{k}+\mib{q}_{\text{tot}}$
    which are on the same Fermi surface.
    This pair can be scattered to another Fermi surface.
  }
\end{figure}

Among possible multi-Fermi-surface cases,
a system with pocket Fermi surfaces, as shown in Fig.~\ref{figure:pocket_FS},
is interesting for superconductivity.
In such a system, a pair on a Fermi surface has
a finite total momentum $\mib{q}_{\text{tot}}$ like
the Fulde-Ferrell-Larkin-Ovchinnikov (FFLO) state~\cite{Fulde,Larkin}
even without a magnetic field.
When the electrons comprised in this pair are scattered to another Fermi surface
by fluctuations such as spin fluctuations and can form a pair on that Fermi surface,
such a pairing state with finite $\mib{q}_{\text{tot}}$ can be stabilized.

In this Letter, in order to explicitly discuss the possibility
of such exotic superconductivity,
we consider an $e_g$ orbital model on a square lattice as an example.
This model can easily realize pocket Fermi surfaces
even if we consider only nearest-neighbor hopping.
To investigate the possible superconducting states in this model,
we apply FLEX approximation that has been extended
to multi-orbital models~\cite{Kubo_2OHM,Takimoto2004,Mochizuki,Mochizuki2,Yada,Kubo_Ce115}.
Then, we show that such pairing states with a finite total momentum
like the FFLO state are stabilized in a system with pocket Fermi surfaces.

We consider a tight-binding model for $e_g$ orbitals given by
\begin{equation}
  \begin{split}
    H=&\sum_{\mib{k},\tau,\tau^{\prime},\sigma}
    \epsilon_{\mib{k} \tau \tau^{\prime}}
    c^{\dagger}_{\mib{k} \tau \sigma}c_{\mib{k} \tau^{\prime} \sigma}
    +U \sum_{i, \tau}
    n_{i \tau \uparrow} n_{i \tau \downarrow}\\
    &+U^{\prime} \sum_{i}
    n_{i 1} n_{i 2}
    + J \sum_{i,\sigma,\sigma^{\prime}}
    c^{\dagger}_{i 1 \sigma}
    c^{\dagger}_{i 2 \sigma^{\prime}}
    c_{i 1 \sigma^{\prime}}
    c_{i 2 \sigma}
    \\
    &+ J^{\prime}\sum_{i,\tau \ne \tau^{\prime}}
    c^{\dagger}_{i \tau \uparrow}
    c^{\dagger}_{i \tau \downarrow}
    c_{i \tau^{\prime} \downarrow}
    c_{i \tau^{\prime} \uparrow},
  \end{split}
  \label{eq:H}
\end{equation}
where $c_{i\tau\sigma}$ is the annihilation operator of
the electron at site $i$ with orbital $\tau$ ($=1$ or 2)
and spin $\sigma$ ($=\uparrow$ or $\downarrow$),
$c_{\mib{k}\tau\sigma}$ is its Fourier transform,
$n_{i \tau \sigma}=c^{\dagger}_{i \tau \sigma} c_{i \tau \sigma}$, and
$n_{i \tau}=\sum_{\sigma}n_{i \tau \sigma}$.
The coupling constants $U$, $U^{\prime}$, $J$, and $J^{\prime}$
denote the intra-orbital Coulomb, inter-orbital Coulomb, exchange,
and pair-hopping interactions, respectively.
In this study, we use the relations
$U=U^{\prime}+J+J^{\prime}$ and $J=J^{\prime}$~\cite{Tang}.

We consider the hopping integrals for the nearest-neighbor sites on a square lattice,
and the coefficients of the kinetic energy terms in Eq.~\eqref{eq:H}
are generally given by
$\epsilon_{\mib{k} 11}=\frac{1}{2}[3(dd\sigma)+(dd\delta)](\cos k_x+\cos k_y)$,
$\epsilon_{\mib{k} 22}=\frac{1}{2}[(dd\sigma)+3(dd\delta)](\cos k_x+\cos k_y)$, and
$\epsilon_{\mib{k} 12}=\epsilon_{\mib{k} 21}=-\frac{\sqrt{3}}{2}[(dd\sigma)-(dd\delta)](\cos k_x-\cos k_y)$,
where $(dd\sigma)$ and $(dd\delta)$ are Slater-Koster integrals,
and the lattice constant is set to unity~\cite{note1}.

To realize pocket Fermi surfaces in a system
such as that shown in Fig.~\ref{figure:pocket_FS},
we set $(dd\sigma)=-(dd\delta)=\frac{2}{\sqrt{3}}t$.
\begin{figure}[t]
  \begin{center}
    \includegraphics[width=8.5cm]{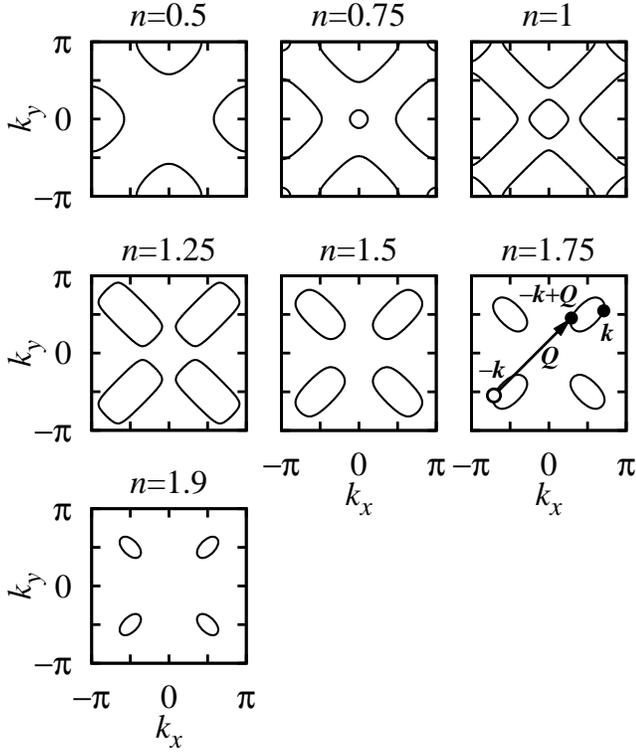}
  \end{center}
  \caption{\label{figure:FS}
    Fermi surfaces of the model for $(dd\sigma)=-(dd\delta)$
    for several electron numbers $n$ per site.
}
\end{figure}
Figure~\ref{figure:FS} shows the Fermi surfaces for
several electron numbers $n$ per site.
Due to the electron-hole symmetry, it is enough to consider $0 \le n \le 2$.
For $n=2$, the Fermi surfaces disappear in this model.
Pocket Fermi surfaces are realized for $n \gtrsim 1.25$.
In this system,
if an electron with $-\mib{k}$ is on a Fermi surface,
the electron with $-\mib{k}+\mib{Q}$ is on another Fermi surface
where $\mib{Q}=(\pi,\pi)$,
as shown in Fig.~\ref{figure:FS} for $n=1.75$ as an example.
Then, it is possible to form a pair
with the total momentum $\mib{q}_{\text{tot}}=\mib{Q}$
by electrons with $\mib{k}$ and $-\mib{k}+\mib{Q}$.
Thus, in this study,
we consider superconducting states with $\mib{q}_{\text{tot}}=\mib{Q}$
in addition to the ordinary superconducting states with $\mib{q}_{\text{tot}}=(0,0)$.

The Green's function for the present two-orbital model
is expressed by a $2 \times 2$ matrix.
In the normal phase,
the Dyson-Gorkov equation
for the Green's function in matrix form is given by
\begin{equation}
  G(k)=G^{(0)}(k)+G^{(0)}(k)\Sigma(k)G(k),
  \label{eq:DG_normal}
\end{equation}
where $k=(\mib{k},\text{i}\epsilon_n)$
and $\epsilon_n=(2n+1)\pi T$ is the Matsubara frequency for fermions
with an integer $n$ and a temperature $T$.
The non-interacting Green's function is given by
$G^{(0)}(\mib{k},\text{i}\epsilon_n)
=[\text{i}\epsilon_n-\epsilon_{\mib{k}}+\mu]^{-1}$,
where $\mu$ is the chemical potential.
In the FLEX approximation, the self-energy is given by
\begin{equation}
  \Sigma_{\tau \tau^{\prime}}(k)=\frac{T}{N}\sum_{q, \tau_1, \tau_2}
  V_{\tau \tau_1;\tau^{\prime} \tau_2}(q)G_{\tau_1 \tau_2}(k-q),
\end{equation}
where $N$ is the number of lattice sites,
$q=(\mib{q},\text{i}\omega_m)$, and $\omega_m=2m\pi T$
is the Matsubara frequency for bosons with an integer $m$.
The matrix $V(q)$ is written as
\begin{equation}
  \begin{split}
    V(q)
    =&\frac{3}{2}[U^{\text{s}} \chi^{\text{s}}(q) U^{\text{s}}
    -U^{\text{s}} \chi^{(0)}(q) U^{\text{s}}/2
    +U^{\text{s}}]
    \\
    +&\frac{1}{2}[U^{\text{c}} \chi^{\text{c}}(q) U^{\text{c}}
    -U^{\text{c}} \chi^{(0)}(q) U^{\text{c}}/2
    -U^{\text{c}}],
  \end{split}
  \label{eq:V_normal}
\end{equation}
where the matrix elements of $U^{\text{s}}$ and $U^{\text{c}}$ are given by
$U^{\text{s}}_{11;11}=U^{\text{s}}_{22;22}
=U^{\text{c}}_{11;11}=U^{\text{c}}_{22;22}=U$,
$U^{\text{s}}_{11;22}=U^{\text{s}}_{22;11}=J$,
$U^{\text{c}}_{11;22}=U^{\text{c}}_{22;11}=2U^{\prime}-J$,
$U^{\text{s}}_{12;12}=U^{\text{s}}_{21;21}=U^{\prime}$,
$U^{\text{c}}_{12;12}=U^{\text{c}}_{21;21}=-U^{\prime}+2J$,
$U^{\text{s}}_{12;21}=U^{\text{s}}_{21;12}
=U^{\text{c}}_{12;21}=U^{\text{c}}_{21;12}=J^{\prime}$;
the other matrix elements are zero.
The susceptibilities $\chi^{\text{s}}(q)$ and $\chi^{\text{c}}(q)$ are given by
$\chi^{\text{s}}(q)=\chi^{(0)}(q)[1-U^{\text{s}} \chi^{(0)}(q)]^{-1}$ and
$\chi^{\text{c}}(q)=\chi^{(0)}(q)[1+U^{\text{c}} \chi^{(0)}(q)]^{-1}$,
respectively, in matrix forms.
The matrix elements of $\chi^{(0)}(q)$ are defined by
$\chi^{(0)}_{\tau_1 \tau_2; \tau_3 \tau_4}(q)
=-\frac{T}{N}\sum_{k}G_{\tau_1 \tau_3}(k+q)G_{\tau_4 \tau_2}(k)$.
%
We solve Eqs.~\eqref{eq:DG_normal}--\eqref{eq:V_normal} self-consistently.
Then, we can calculate response functions; for example,
the spin susceptibility in the FLEX approximation is given by
$
  \chi(q)
  =2\sum_{\tau,\tau^{\prime}}
  \chi^{\text{s}}_{\tau \tau; \tau^{\prime} \tau^{\prime}}(q)
$.
  
The linearized gap equation for the anomalous self-energy
$\phi^{\xi}_{\tau \tau^{\prime}}(k;\mib{q}_{\text{tot}})$ is given by
\begin{equation}
  \begin{split}
    \phi^{\xi}_{\tau \tau^{\prime}}&(k;\mib{q}_{\text{tot}})
    =-\frac{T}{N}\sum_{k^{\prime},\tau_1,\tau_2,\tau_3,\tau_4}
    V^{\xi}_{\tau \tau_1; \tau_2 \tau^{\prime}}(k-k^{\prime})\\
    \times& G_{\tau_1 \tau_3}(k^{\prime})
    \phi^{\xi}_{\tau_3 \tau_4}(k^{\prime};\mib{q}_{\text{tot}})
    G_{\tau_2 \tau_4}(-\mib{k}^{\prime}+\mib{q}_{\text{tot}},-\text{i}\epsilon_n),
  \end{split}
  \label{eq:gap_eq}
\end{equation}
where the spin state is denoted by $\xi=\text{singlet}$ or triplet,
and $\mib{q}_{\text{tot}}$ denotes the total momentum of the pair.
The superconducting transition temperature is given by the temperature
for which Eq.~\eqref{eq:gap_eq} has a nontrivial solution.
The effective pairing interactions $V^{\xi}(q)$ are written as
\begin{align}
  V^{\text{singlet}}(q)=[3&V^{\text{s}}(q)-V^{\text{c}}(q)]/2,
  \label{eq:V_singlet}
  \\
  V^{\text{triplet}}(q)=[-&V^{\text{s}}(q)-V^{\text{c}}(q)]/2,
  \label{eq:V_triplet}
\end{align}
where
$V^{\text{s}}(q)=[U^{\text{s}} \chi^{\text{s}}(q) U^{\text{s}}+U^{\text{s}}/2]$
and
$V^{\text{c}}(q)=[U^{\text{c}} \chi^{\text{c}}(q) U^{\text{c}}-U^{\text{c}}/2]$.

Before presenting the calculation results,
we discuss plausible candidates for the pairing symmetry
from Eqs.~\eqref{eq:gap_eq}--\eqref{eq:V_triplet}.
First, in this discussion, we ignore the orbital indices in the above equations for simplicity.
In this model, we find that only the spin susceptibility becomes large.
Then, we expect an anisotropic superconducting state for the spin-singlet channel,
since $V^{\text{singlet}}(q)$ becomes large and has a positive value
and the anomalous self-energy has to change its sign in the $\mib{k}$ space,
as understood from Eq.~\eqref{eq:gap_eq}.
For the spin-triplet channel,
an $s$-wave state is favorable,
since $V^{\text{triplet}}(q)$ becomes negative
and the anomalous self-energy does not have to change its sign.
The above consideration is appropriate for orbital-parallel components, i.e.,
$\phi^{\xi}_{11}(k;\mib{q}_{\text{tot}})$ and $\phi^{\xi}_{22}(k;\mib{q}_{\text{tot}})$,
as long as the Pauli principle is satisfied.
On the other hand, for orbital-antiparallel components
$\phi^{\xi}_{12}(k;\mib{q}_{\text{tot}})$ and $\phi^{\xi}_{21}(k;\mib{q}_{\text{tot}})$,
orbital indices transform following the $d_{x^2-y^2}$ symmetry
under symmetry operations of the lattice,
and thus,
the anomalous self-energy belonging to the $d_{x^2-y^2}$ symmetry
does not have to change its sign in the $\mib{k}$ space.
In other words, the symmetry in the $\mib{k}$ space
of a $d_{x^2-y^2}$-wave state for the orbital-antiparallel components is the same as that
of an $s$-wave state for the orbital-parallel components.
Thus, spin-triplet $d_{x^2-y^2}$-wave states
and spin-singlet states other than those with the $d_{x^2-y^2}$ symmetry
are favorable for the orbital-antiparallel components.
Here, we note that orbital states mix in general.
However, it is expected that
when the Hund's rule coupling $J$ is large
and the inter-orbital Coulomb interaction $U^{\prime}$ is small,
the orbital-antiparallel components become dominant.


%
\begin{figure}[t]
  \begin{center}
    \includegraphics[width=7.2cm]{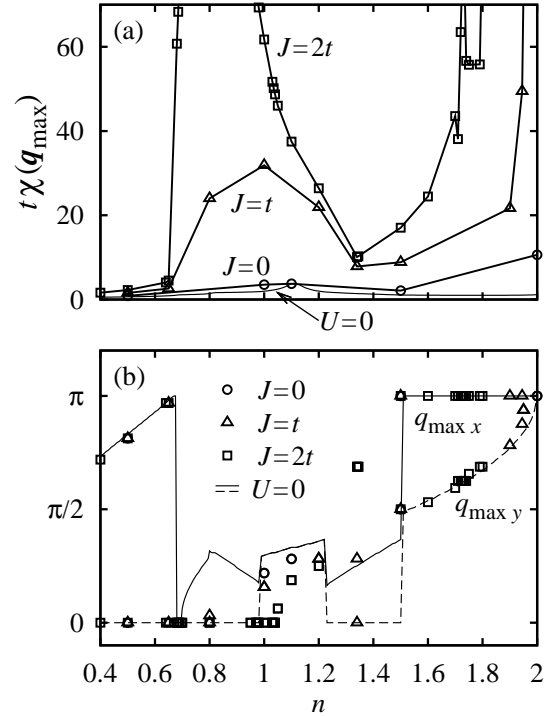}
  \end{center}
  \caption{\label{figure:max_q}
    (a) spin susceptibility $\chi(\mib{q})$ at $\mib{q}=\mib{q}_{\text{max}}$
    and (b) $\mib{q}_{\text{max}}$ for $J=0$ (circles), $t$ (triangles),
    and $2t$ (squares)
    as functions of $n$ for $U=6t$ at $T=0.005t$.
    The thin lines are for $U=0$.
  }
\end{figure}

Here,
we show the results for a $64 \times 64$ lattice.
In this study, we fix the value of
the intra-orbital Coulomb interaction $U=6t$
and vary $J$ ($=J^{\prime}$).
Then, the inter-orbital Coulomb interaction is given by
$U^{\prime}=U-2J$.
In Fig.~\ref{figure:max_q}(a),
we show the $n$ dependence of
the static spin susceptibility $\chi(\mib{q})=\chi(\mib{q},\text{i}\omega_m=0)$
at $\mib{q}=\mib{q}_{\text{max}}$ for $J=0$, $t$, and $2t$ at $T=0.005t$,
where $\mib{q}_{\text{max}}$ is defined as the wave vector
at which $\chi(\mib{q})$ becomes the largest.
For comparison purposes, we also show $\chi(\mib{q}_{\text{max}})$
for the non-interacting system.
The spin susceptibility is enhanced by the interactions,
in particular, by the Hund's rule coupling.
On the other hand, $\mib{q}_{\text{max}}$ does not depend so much on
the interactions as shown in Fig.~\ref{figure:max_q}(b).
This fact indicates that
the characteristic wave vector $\mib{q}_{\text{max}}$
is almost determined by the property of the non-interacting system, i.e.,
the Fermi-surface structure.

As shown in Fig.~\ref{figure:max_q}(b),
the characteristic wave vector $\mib{q}_{\text{max}}$ takes various values
depending on $n$ around $n=0.7$--1.5,
since the Fermi surface structure changes around this area,
as shown in Fig.~\ref{figure:FS},
e.g, the Fermi surfaces consist of electron sheets for $n=1$,
while they consist of hole pockets for $n=1.25$.
Thus, several fluctuations exist in this region,
and a $d_{x^2-y^2}$-wave spin-triplet state with dominant orbital-antiparallel
components may appear
if such fluctuations in a wide $\mib{q}$ region are available.
However, around $n=0.7$--1.1, the spin fluctuations are very strong for $J=2t$
and thus magnetic order should occur.
For $n\gtrsim 1.5$, we expect the appearance of a $p$-wave spin-singlet state
because of the following reason.
\begin{figure}[t]
  \begin{center}
    \includegraphics[width=0.7\linewidth]{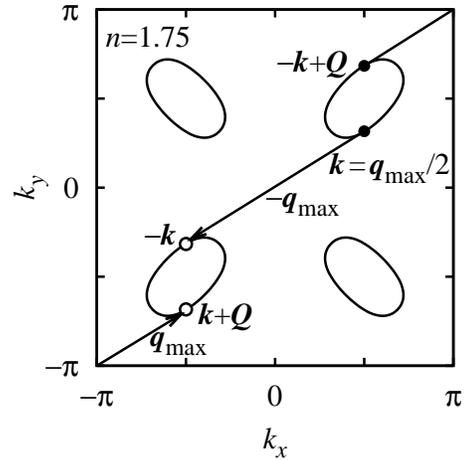}
  \end{center}
  \caption{\label{figure:FS_N1.75}
    Fermi surfaces for $n=1.75$.
  }
\end{figure}
For the purpose of explanation, we show the Fermi surfaces for $n=1.75$
in Fig.~\ref{figure:FS_N1.75}.
The characteristic wave vector $\mib{q}_{\text{max}}$ connects
electrons at $\mib{k}=\mib{q}_{\text{max}}/2$ and $-\mib{k}$.
Thus, electrons around $\mib{k}$ and $-\mib{k}+\mib{Q}$ comprising a pair
on a Fermi surface can be scattered to another Fermi surface
by the spin fluctuations and can comprise a pair with momenta around
$-\mib{k}$ and $\mib{k}+\mib{Q}$.
These pairs are the space inversions of each other, and to utilize such fluctuations,
the $p$-wave spin-singlet state is favorable
since the anomalous self-energy changes its sign
under the inversion $\mib{k} \rightarrow -\mib{k}$.
These $d_{x^2-y^2}$-wave spin-triplet and $p$-wave spin-singlet states
should be dominated by orbital-antiparallel components because of
the Pauli principle.
The orbital-parallel components should be odd in frequency
for these superconducting states and play only a minor role.

\begin{figure}[t]
  \begin{center}
    \includegraphics[width=7.8cm]{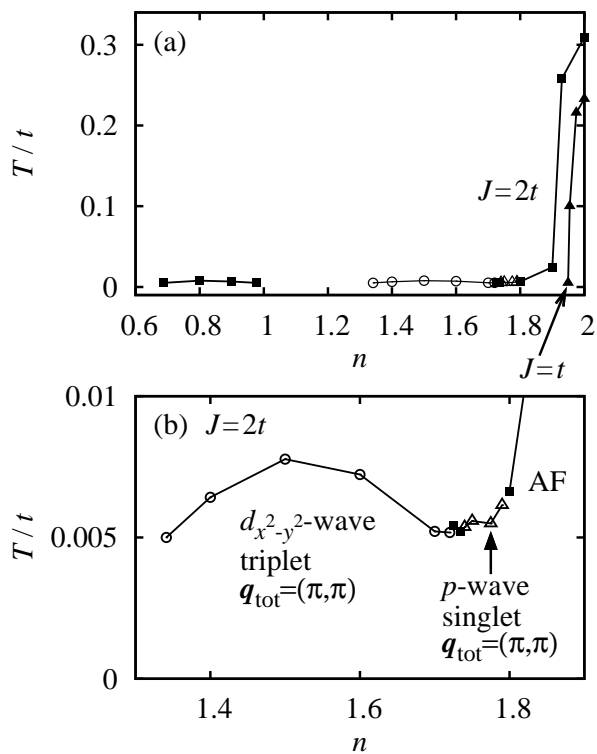}
  \end{center}
  \caption{\label{figure:PD}
    (a) Transition temperatures for $U=6t$.
    The solid triangles and squares
    denote antiferromagnetic (AF) transition temperatures
    for $J=t$ and $2t$, respectively.
    The open circles and triangles denote
    superconducting transition temperatures for
    the $d_{x^2-y^2}$-wave spin-triplet state and
    $p$-wave spin-singlet state,
    respectively, for $J=2t$.
    In both superconducting states,
    Cooper pairs have the total momentum $\mib{Q}=(\pi,\pi)$.
    (b) Transition temperatures around $n=1.6$ for $J=2t$.
  }
\end{figure}
Figure~\ref{figure:PD}(a) shows
the highest transition temperatures
among transition temperatures for all superconducting states allowed in the model
and for antiferromagnetic states
as functions of $n$
for $J=t$ and $2t$.
Figure~\ref{figure:PD}(b) shows the highest transition temperatures
for $J=2t$ around $n=1.6$.
The antiferromagnetic transition temperatures are determined
by $\chi(\mib{q}_{\text{max}})=70/t$,
and the system is always in the normal phase for $J=0$ at $T\ge 0.005t$.
The antiferromagnetic phase extends by increasing the Hund's rule coupling.
With regard to superconductivity,
we cannot find any superconducting phase within $T\ge 0.005t$ for $J=t$.
For $J=2t$, a $p$-wave spin-singlet state with $\mib{q}_{\text{tot}}=\mib{Q}$
appears, as is expected from Fig.~\ref{figure:FS_N1.75},
at $n \simeq 1.75$ where the antiferromagnetic transition temperature tends to zero.
The $d_{x^2-y^2}$-wave spin-triplet state with $\mib{q}_{\text{tot}}=\mib{Q}$
appears in a wider region $n\simeq 1.3$--1.7,
since the $d_{x^2-y^2}$-wave spin-triplet state is
insensitive to the value of the characteristic wave vector $\mib{q}_{\text{max}}$,
similar to the $s$-wave state in the two-orbital Hubbard model~\cite{Kubo_2OHM}.
It should be noted that these superconducting states with finite $\mib{q}_{\text{tot}}$
appear only in the region where the system has pocket Fermi surfaces.

To summarize,
we have pointed out that a pairing state with a finite total momentum
like the FFLO state is possible for a system with pocket Fermi surfaces.
As an example, we have studied a model for $e_g$ orbitals
by applying FLEX approximation.
Then, we have found that
the $p$-wave spin-singlet and $d_{x^2-y^2}$-wave spin-triplet states
with $\mib{q}_{\text{tot}}=\mib{Q}=(\pi,\pi)$ are realized
at electron number $n$ where the system has pocket Fermi surfaces.
The mechanism to stabilize the superconducting states
with finite $\mib{q}_{\text{tot}}$ can be applied
to systems other than the $e_g$ orbital system
as long as such pocket Fermi surfaces exist.

The author thanks T. Hotta, T. Maehira, and T. D. Matsuda for useful comments.
This work is supported by Grants-in-Aid for Scientific Research
in Priority Area ``Skutterudites''
and for Young Scientists
from the Ministry of Education, Culture, Sports, Science
and Technology of Japan.



\begin{thebibliography}{99}
\bibitem{note1}
  This model can also be regarded as a model for the $\Gamma_8$ orbitals
  of $f$ electrons for the total angular momentum $j=5/2$
  by replacing
  $(dd\sigma)$ by $1008[3(ff\sigma)+4(ff\pi)]$ and
  $(dd\delta)$ by $336[(ff\pi)+5(ff\delta)+15(ff\phi)]$.
\end{thebibliography}
\end{document}